\documentclass[runningheads]{llncs}

\usepackage{abbrevs}
\usepackage{amsmath} 
\usepackage{amssymb} 
\usepackage[english]{babel}
\usepackage{balance}
\usepackage{booktabs}
\usepackage{epsfig} 
\usepackage{graphicx}
\usepackage[utf8]{inputenc}
\usepackage{mathptmx} 
\usepackage{mdframed}
\usepackage{pifont}
\usepackage{times}
\usepackage{tikz}
\usepackage{xcolor}
\usetikzlibrary{arrows,automata}
\usepackage{url}

\usepackage{breakurl}
\usepackage[breaklinks]{hyperref}
\usepackage[normalem]{ulem}

\usepackage{comment}
\usepackage{xspace}
\usepackage{stmaryrd}

\DeclareMathAlphabet{\mathcal}{OMS}{cmsy}{m}{n}

\newcommand{\todo}[1]{\textcolor{red}{TODO: #1}}

\newcommand{\ea}[1]{\textcolor{blue}{E.A.: #1}}
\newcommand{\ad}[1]{\textcolor{brown}{A.D.: #1}}
\newcommand{\rl}[1]{\textcolor{purple}{R.L.: #1}}
\renewcommand{\ea}[1]{} 
\renewcommand{\ad}[1]{} 
\renewcommand{\rl}[1]{} 
\renewcommand{\todo}[1]{} 
\newcommand{\proba}[1]{\mathbb{P}[#1]}

\newcommand{\nom}{\textsc{StakeCube}\xspace}
\newcommand{\hash}{\mathfrak{h}}
\newcommand{\Smin}{s_{\min}}
\newcommand{\Smax}{s_{\max}}

\begin{document}

\title{
\nom: Combining Sharding and Proof-of-Stake to build Fork-free Secure Permissionless Distributed Ledgers
}

\author{Antoine Durand\inst{1}
\and Emmanuelle Anceaume\inst{2}
\and Romaric Ludinard\inst{3}
}

\institute{
IRT SystemX, Paris-Saclay, France \email{antoine.durand@irt-systemx.fr}
\and
CNRS, Univ Rennes, Inria, IRISA France \email{emmanuelle.anceaume@irisa.fr} \and
IMT Atlantique, IRISA France \email{romaric.ludinard@imt-atlantique.fr}
}

\pagestyle{plain}

\maketitle

\begin{abstract}
Our work focuses on the design of a scalable permissionless blockchain in the proof-of-stake setting. In particular, we use a distributed hash table as a building block to set up randomized shards, and then leverage the sharded architecture to validate blocks in an efficient manner. We combine verifiable Byzantine agreements run by shards of stakeholders and a block validation protocol to guarantee that forks occur with negligible probability. We impose induced churn to make shards robust to eclipse attacks, and we rely on the UTXO coin model to guarantee that any stakeholder action is securely verifiable by anyone.  Our protocol works against adaptive adversary, and makes no synchrony assumption beyond what is required for the byzantine agreement.

\end{abstract}


\section{Introduction}
\label{sec:introduction}

Permissionless blockchains, also called distributed ledgers, initially appeared as the technological solution for the deployment of the Bitcoin digital cryptocurrency and payment system~\cite{nakamoto2008bitcoin}. 
Permissionless blockchains aim at achieving the impressive result of being a persistent, distributed, consistent and continuously growing log of transactions, publicly auditable and writable by anyone. Despite the openness of the environment and thus the inescapable presence of malicious behaviors, security and consistency of permissionless blockchains do not demand the presence of a trusted third party.

This is a real achievement, which mainly results from the tight combination of two ingredients: a randomized election of the next block of transactions to be appended to the blockchain and a short latency broadcast primitive. While the latter one relies on the properties of peer-to-peer networks, the former one has so far been commonly implemented by solving proof-of-work (PoW), a cryptographic puzzle that is provably secure against a large proportion of participants that may wish to disrupt the system, and allows to keep the rate at which blocks are created parametrizable and independent of the size of the system. This second aspect is important to guarantee that the ratio between the message transmission delay and the block time interval remains low enough whatever the system activity, guaranteeing accordingly an easy management of conflicting blocks, if any.

Unfortunately, resilience of PoW-based solutions fundamentally relies on the massive use of computational resources, which is a real issue today. Lot of investigations have been devoted to find a secure alternative to PoW, but most of them either rely on the intensive use of a large quantity of physical resources (\eg, proof-of-space~\cite{PoSpace}, proof-of-space/time~\cite{PoSpace-Time}) or makes compromises in their trust assumptions (\eg proof-of-elapsed-time~\cite{poet}, delegated proof-of-stake~\cite{eos}). In contrast, solutions based on proof-of-stake (PoS) seem to be a quite promising way to build secure and permissionless blockchains. Indeed, proof-of-stake rely on a limited but abstract resource, the crypto-currency, in such a way that the probability for a participant to create the next block of the blockchain is generally proportional to the fraction of currency owned by this participant. It is an elegant alternative in the sense that all the information needed to verify the legitimacy of a stakeholder to create a block (i.e., crypto-currency possession) is already stored in the blockchain. Finally, by being a sustainable alternative (creating a block requires a few number of operations), scalability concerns, exhibited by PoW-based solutions, should be a priori more tractable.

An important condition for a PoS-blockchain to be secure is randomness. The creator of the next block must be truly random, and the source of randomness must not be biaised by any adversarial strategy. So far, this has been achieved by two main approaches: chain-based consensus and block-wise Byzantine agreement with respectively Ourobouros~\cite{OuroborosPraos} and Algorand~\cite{algorand} as main representatives. In the former approach, a snapshot of the current users' status is periodically taken, from which the the next sequence of leaders is computed. In the latter one, a Byzantine agreement per block, relying on the properties of verifiable random cryptographic schemes, is achieved. High robustness against adaptive adversarial strategies results from the dynamic participation of thousands of users, each one participating for a single step of the algorithm.

In this paper we present a new blockchain protocol called \nom which aims at improving scalability of the block-wise Byzantine agreement approach by combining sharding techniques, users presence and stake transfer to operate in a PoS setting. 
The key idea of \nom is to organise users (\ie stakeholders) into shards--- such that the number of shards increases sub-linearly with the total number of active UTXOs--- and within each shard, to randomly choose a constant size committee in charge of executing the distributed algorithms that contribute to the creation of blocks. 
Each block at height   $h$ in the blockchain  is by design  unique (no fork), and once a block is accepted in the blockchain, the next one is created by a sub-committee of shards whose selection is random with a distribution that depends on the content of  the last accepted block. 

To make such a solution correct in presence of a Byzantine adversary,
we  guarantee that the adversary cannot predict the shards in which users will sit, and that the sojourn time of users in their shard is limited. Doing so is an effective way to protect the system against eclipse attacks~\cite{ASLT11,AS07}. We introduce the notion of unpredictable and perishable users' credentials. Then to cope with this induced churn, shards' views are updated, signed  and installed once, and this occurs right before the acceptation of a new block. Finally, the creation of blocks  is efficiently handled by an agreement among a  verifiable sub-committee of shards. We might expect that solely relying on stakeholders (\ie, owners of the coins of the crypto-currency system) to the secure construction of the blockchain makes sense due to their incentive to be fully involved in the blockchain governance, rather than delegating it to powerful miners. However the analysis against rational players is left as future work.

The remaining of the paper organised as follows.
Section~\ref{sec:related-work} presents related work, Section~\ref{sec:model} details our model and assumptions while Section~\ref{sec:addressed-problem} formalises  the addressed  problem. Section~\ref{sec:set-of-ingredients} describes  an high-level view of the required building blocks of \nom  while Section~\ref{sec:validation-protocols} presents the design principles of the proposed solution. A security analysis is provided in Section~\ref{sec:security} before concluding in Section~\ref{sec:conclusion}.

\section{Related work}
\label{sec:related-work}

Omniledger~\cite{Omniledger} is the closest work to ours. It is a  PoS-compatible, sharded,  distributed ledger, resilient against a weakly dynamic adversary that corrupts up to $\frac{1}{4}$ of participants. In contrast to our approach, Omniledger assumes a strongly synchronous setting, and each shard maintains its own ledger and, global synchronisation of transactions is achieved through  an atomic commit protocol tailored to their usage. 
Ouroboros~\cite{Ouroboros}, representative of the chain-based approach, is a synchronous PoS protocol resilient against a weakly dynamic adversary that owns $1/2-\epsilon$ of stake. Moreover, Ouroboros has been recently improved to work in the partially synchronous setting against a dynamic adversary~\cite{OuroborosPraos,OuroborosGenesis}, but keeping the same design principles as the original one. In Ouroboros, a unique leader is elected at each round to broadcast its block which contrasts with our sharded approach where the block creation process is distributed.
Snow White~\cite{SnowWhite} is a synchronous PoS protocol resilient against a weakly dynamic adversary that owns $1/2$ of the \emph{active} stake. This protocol also relies on a leader election.
Algorand~\cite{algorand}, is a representative of the blockwise Byzantine agreement approach. It provides a distributed ledger against an strongly adaptive adversary without assuming strong synchrony assumptions. However, by its design, agreement for each block of the blockchain is achieved by involving a very large number of stakeholders  so that each one needs to effectively participate only for one exchange of messages.

\section{Model}
\label{sec:model}
We assume a large, finite set of users whose composition may change over time. Users do not
have synchronized clocks, but their individual clocks drift at the same rate. Users communicate by propagating messages within the system. The delivery of network messages is at the discretion of the adversary, but subject to synchrony assumptions.
Our construction in itself makes no synchrony assumption except for what is required for the Byzantine resilient building blocks.
Since our construction uses multiple building blocks, synchrony assumptions may be changed if they are instantiated differently than suggested.
Users have access to basic cryptographic functions, including a cryptographic hash function $\hash$, and a CPA-secure signature scheme. Function $\hash$ is modeled as a random oracle.
Users own some minimal amount of stake (\ie money), which gives them the right to participate to \nom. We adopt (a simplified version of) what is commonly known as the Bitcoin Unspent Transaction Output (UTXO) model. An UTXO can be roughly seen as a user's account credited by some stake. 
An UTXO is uniquely characterized by a public key $pk_i$ and its associated amount of stake $s_i$. Each public key is related to the digital signature schema \(\Sigma\) with the uniqueness property, which allows stakeholders to use the public keys (or a hash thereof) of their UTXOs as a reference to them, as demonstrated in the "Public Keys as Identities principle" of Chaum~\cite{chaum88}. 
Note that the number of users evolves according to the UTXO set.
At any time, a user can own multiple UTXOs. UTXOs can be debited only once, and once debited, an UTXO does not exist anymore. 
To simplify discussion, transactions outputs do not contain $\hash(pk_i)$ but directly $pk_i$.
\paragraph{Threat model: A weakly adaptive adversary}
We assume the presence of Byzantine (\ie malicious) users which controls up to $\mu \leq  1/3-\epsilon$ of the total amount of stake currently available in the system. Here, $\epsilon$ quantifies the gain in the effective adversarial power, related to the security parameter. 
This model, named the "Stake Threshold Adversary" by Abraham and Malkhi~\cite{AM2017}, is an alternative to the common Threshold Adversary Model, which bounds the total number of parties the adversary controls relative to the total population of the system, and an extension (or modification) of the Computational Threshold Adversary introduced by Bitcoin, which bounds the proportion of the computational power owned by parties. Byzantine users can deviate from the protocol. They are modeled by an adversary. The adversary can perfectly coordinates all  malicious users. It can learn the messages sent by honest users (\ie non malicious users), delay them, and then chooses messages sent by malicious ones. Further the adversary is weakly adaptive: it can select at any time which users to corrupt in replacement of corrupted ones (\ie corruptions are "moving"), however a corruption becomes effective $T$ blocks after the adversary has selected the user to be corrupted. 
The adversary is computationally bounded so that it can neither forge honest nodes' signatures nor break the hash function and the signature scheme. 
Finally, we  assume that all users (honest and malicious) share an initial knowledge that we call \emph{genesis block} which contains an initial arbitrary UTXO set. We assume this block also shares the same properties as regular blocks. How to setup the genesis block is out of the scope of this paper.

\section{The Addressed Problem} 
\label{sec:addressed-problem}

\nom aims at allowing any honest user $i$  to locally maintain a sequence of blocks $B_0^i, B_1^i, \ldots, B_h^i$, where $h$ represents the index (or the height) of the block in the sequence. This sequence of blocks represents  $i$'s copy of the distributed ledger, and satisfies both  Safety and  Liveness properties. In addition, the orchestration of the shards allows \nom to satisfy both Scalability and Efficiency properties. \nom is parametrized with an arbitrary security parameter $\kappa$, so  that all its properties are guaranteed with probability at least $1-e^{-O(\kappa)}$.

\begin{property}[\textbf{Safety}]
If honest user $i$ accepts a block $B_h^i$ at height $h$ in its copy of the ledger then, for any honest user $j$ that accepts a block at height  $h$ in its copy ledger, $B_h^j=B_h^i$.
\end{property}

\begin{property}[\textbf{Liveness}]
If a honest user submits transaction $tx$, then eventually $tx$ appears in a block accepted in the copy of all  honest users.
\end{property}

In \nom, participation of honest users is conditional to the possession of UTXOs. Participation is voluntary: Any honest user can join a  shard (determined by the protocol), whenever she wishes, with the objective of eventually being involved in the Byzantine resilient protocols executed in this shard. Participation is temporary: The sojourn time of an honest user in a shard is defined by the time it takes for \nom to create $T$ blocks. Once she leaves, she can participate again by joining another shard, and does so until she spends her UTXO.
As users may own multiple UTXOs, they can simultaneously and verifiably sit in different shards. 
In the following, a user that issues a join request with its current credential is called an \emph{active} user. 
\noindent
\nom satisfies Scalability and Efficiency properties. This is achieved due to the properties of the block creation process. Adding a new block takes two Byzantine fault tolerant protocols to be run in parallel within each shard, one network wide diffusion by each shard, one inter-shard byzantine agreement, and finally one broadcast for the block (more details will be given in Section~\ref{sec:construction-next-block}).
\begin{property}[\textbf{Scalability}]
All Byzantine fault tolerant protocols we rely on have an (overall) $O(n^3)$ message complexity. However in \nom these protocols are executed by committees whose size is small and fixed.  Because the number of shard is $O(\sqrt{N})$, the overall communication cost is $O(N C_1^3+C_2^3)$, with $C_1$ and $C_2$ some constants depending on $\kappa$. 
Thus, each participant's average communication cost is sublinear in $N$.
\end{property}

\begin{property}[\textbf{Efficiency}] All Byzantine fault tolerant protocols we rely on use a constant number of rounds. Thus adding a new block also takes a constant number of rounds. Because a transaction, once diffused, will be included in the next block and blocks are permanently attach to the blockchain, it takes at most two blocks to include a newly received transaction.

\end{property}

\section{A Set of Ingredients}
\label{sec:set-of-ingredients}

To solve the addressed problem, \nom relies on the orchestration of the following ingredients.

\vspace{0.2cm}
\noindent
\textbf{Cryptographic Primitives} 
Digital signature together with random hash functions allow the implementation of verifiable random functions (VRF)~\cite{MRV99}. In a VRF, a secret key $sk$ allows the evaluation $y$ of hash function $\mathfrak{h}$ on input $x$ as well as the computation of a non-interactive proof that shows that the secret key $sk$ is the only one that can compute $y$. Verification of the proof is done with respect to the public key $pk$ only. The proof must remain sound even when $pk$ is computed maliciously and $\mathfrak{h}(sk,x)$ must remain pseudorandom even when an adversary can query values of $\mathfrak{h}$ and proofs for them for any input value $x'$.

\vspace{0.2cm}
\noindent
\textbf{Byzantine Vector Consensus}
A vector consensus protocol~\cite{vectorconsensus} is a Byzantine resilient protocol where $n$ participants agree on a vector representing the input value of each participant. Validity condition states that in presence of \(f\leq\lfloor (n-1)/3\rfloor\) Byzantine nodes, the vector contains at least $f+1$ non-null values, and for each non-null value \(v_i \neq \bot, 1\leq i \leq n\), this value was initially proposed by participant \(i\).

\vspace{0.2cm}
\noindent
\textbf{Random Beacon}
A Random beacon is a service that provides a public source of randomness. It was first proposed by Rabin~\cite{randombeacon83} in the context of contract signing. In our case, we need the random beacon to be emulated by a distributed protocol without trusted third parties, that is, a protocol that satisfies the following security properties:
\begin{enumerate}
    \item \emph{Guaranteed output delivery.} All honest participants eventually output a value.
    \item \emph{Unpredictable.} Any adversary’s ability to predict any information about the beacon prior to it being published is negligible. 
    \item \emph{Unbiased.} For all adversarial strategies, the output is statistically close to a uniformly random string.
    \item \emph{Publicly verifiable.} The protocol also produces a proof that can be verified by third parties to be convinced that a beacon is indeed the output of the protocol.
\end{enumerate}
Suitable instantiations for the distributed setting includes SCRAPE~\cite{scrape} and RandHerd~\cite{randherd}. In the following we denote by $\mu_{core}$ the minimum of the fractional resiliency of the vector consensus and random beacon protocols.

\vspace{0.2cm}
\noindent
\textbf{Verifiable Byzantine Agreement} 
We use a verifiable Byzantine agreement in order to agree on the next signed  block  despite corrupted shards. Our main requirement for this algorithm is to be optimistic, \ie efficient in the absence of faults. Indeed, the analysis in Section~\ref{sec:security} shows that the probability for a shard to be corrupted exponentially decreases with shards core size. Any verifiable Byzantine algorithm satisfying our assumptions can be used. We rely on the solution proposed by Shen et al~\cite{CGMV2018} since it is leader-based, efficient and tolerant to temporary partitions.
The fractional resiliency of this protocol is noted $\mu_{corrupted}$.

\vspace{0.2cm}
\noindent
\textbf{Distributed Hash Table (DHT)} 
Distributed hash tables (DHTs) build their topology according to structured graphs, and for most of them, the following principles hold: each node of the system has an assigned identifier, and the identifier space, \eg, the set of 256-bit strings, is partitioned among all the nodes of the system. Nodes self-organize within the graph according to a distance function  based on the identifier space. 

\vspace{0.2cm}
\noindent
\textbf{Sharded DHT} 
The notion of Sharded DHT is similar to a regular DHT, except that each vertex of the DHT is a set of nodes instead of a single node. That is, nodes gather together into shards, and shards self-organize into a DHT graph topology. Sharded DHTs can be made robust to adversarial strategies as achieved in SChord~\cite{robustchord2005}, and PeerCube~\cite{anceaume2008peercube}, and robust to high churn as achieved in  PeerCube~\cite{anceaume2008peercube} by running Byzantine tolerant algorithms  within each shard. For these reasons,  we rely on  PeerCube architecture, while weakening its model by removing the assumption of a global trusted party supplying verifiable random identifier, and by removing the assumption of a static  adversary. For self-containment  reasons, we now recall the main design features of PeerCube.
Briefly, this is a DHT that conforms to an hypercube. Each vertex (\ie shard) of the hypercube is dynamically formed by gathering nodes that are logically close to each other according to a distance function applied on the identifier space. 
Shards are built so that  the respective common prefix of their members is never a prefix of one-another. This guarantees that each shard has a unique common prefix, that in turn serves as a shard's \emph{label}. The shard's label characterizes the position of the shard in the overall hypercubic topology, as in a regular DHT. 
Shards size is upper and lower bounded. Whenever the size of shard $\mathcal S$ exceeds a given value $\Smax$, $\mathcal S$ splits into two shards such that the label of each of these two new shards is prefixed by $\mathcal S$ label, and whenever the size of $\mathcal S$ falls under a given value  $\Smin$, $\mathcal S$ merges with another shard to give rise to a new shard whose label is a prefix of $\mathcal S$ label. 
Each shard self-organizes into two sets, the core set and the spare set. The core set is a fixed-size random subset of the whole shard. It is responsible for running the Byzantine agreement protocols in order to guarantee that each shard behaves as a single and correct entity (by for example forwarding  all the join and lookup requests to their destination) despite  malicious participants~\cite{anceaume:hal-00736918}. Members of the spare set merely keep track of shard state. Joining the core set only happens when some existing core member leaves, in which case the new member of the core set  is randomly elected among the spare set. By doing this, nodes joining the system weakly impact the topology of the hypercube~\cite{anceaume2008peercube}.
\section{Design Principles of \nom}
\label{sec:validation-protocols}
 \nom allows the creation of a permissionless distributed ledger in a PoS setting. 
The key idea of \nom is to organise users (\ie stakeholders) into shards--- such that the number of shards increases  sub-linearly with the total number of active UTXOs--- and within each shard, to randomly choose a constant size committee in charge of executing the distributed algorithms that contribute to the creation of blocks. 

The randomization of shards members gives a statistical bound the number of malicious participants sitting at each shards, ensuring the correct execution of the agreement primitives. More precisely, we compute bounds that may still cause some shards to have too much malicious participants (\ie they become \emph{corrupted shards}), but the overall number of corrupted shards is bounded. This technique allows us to fix a small shard size while keeping the ability to make security-efficiency trade-offs.

Each block at height $h$ in the blockchain is unique, and is obtained by running an inter-shard agreement procedure among a sub-committee of shards.  

To be able to tolerate the presence of a Byzantine adversary,
we must guarantee that the adversary cannot predict the shards in which users will sit, and that the sojourn time of users in their shard is limited. To achieve this, we introduce the notion of unpredictable and perishable users' credentials in Section~\ref{sec:credential}. Then to cope with this induced churn, we show how to update, sign a install the shards' views in Section~\ref{sec:shard-membership}. This process occurs right before the acceptation of a new block. Finally, as described in Section~\ref{sec:construction-next-block} the creation of blocks is efficiently handled by an agreement among a verifiable sub-committee of shards.
\subsection{Unpredictable and Perishable Users' Credentials}
\label{sec:credential}
As described in Section~\ref{sec:set-of-ingredients}, Peercube critically relies on a (global) trusted party supplying verifiable random identifiers to nodes. In this section, we detail how to construct those in our decentralized setting, using the already known public keys and some randomness present in each block.
For each unspent public key, \ie for each UTXO, owned by a user, a sequence of unpredictable and perishable credentials are tightly assigned to her. Validity of a credential spans $T$ blocks, with $T$ some positive integer. The credential $\sigma$ assigned to user $i$ for its UTXO ($pk_i,sk_i$) is computed as follows. Let $B_{h_0}$ be the  block at height $h_0$ of the blockchain such that $pk_i$ was created in $B_{h_0}$, \ie, it exists a transaction in $B_{h_0}$ such that $pk_i$ appears in the output list of that transaction. For any blockchain height $h\geq h_0+T$,  such that  UTXO ($pk_i,sk_i$) still exists when $B_{h}$ is accepted in the blockchain, 
\begin{align}\label{eq:sojourn_time}
 \sigma_{pk_i}(h) := H( pk_i || B_{h'}.\rho ), & \quad \text{where } \quad h' := h_0 + \lfloor \frac{h-h_0}{T} \rfloor T,
\end{align}
with $B_{h'}.\rho$ a random number whose computation is detailed in Section~\ref{sec:construction-next-block}.
Suppose that $i$'s UTXO ($pk_i,sk_i$) is created in block $B_h$. Then by Relation~\ref{eq:sojourn_time}, $i$'s first credential for UTXO ($pk_i,sk_i$) is computed based on the content of block $B_{h+T}$ and perishes at block $B_{h+2T}$. Then, $i$'s second credential for ($pk_i,sk_i$) is computed based on the content of block $B_{h+2T}$ and perishes at block  $B_{h+3T}$, and so on until $i$ spends ($pk_i,sk_i$). 
User $i$'s credential uniquely characterizes the shard to which user $i$ is allowed to sit, and this shard is the one whose label prefixes $i$'s current credential $\sigma_{pk_i}(h)$. By the non-inclusion property of PeerCube~\cite{anceaume2008peercube}, there does not exist a shard whose label is the prefix of another shard, and thus, there is a unique shard whose label prefixes credential $\sigma_{pk_i}(h)$. When her current credential expires, $i$ leaves the shard she is in, and if she wants to continue to participate to \nom,  joins a new shard based on her new credential. 

\vspace{0.3cm}
\noindent
There are a couple of details that should be noted. 
\begin{enumerate}
    \item User $i$ does not need to participate in \nom for the entire life of her UTXO ($pk_i,sk_i$). She can join \nom (\ie join a shard) at any time $h$ under credential $\sigma_{pk_i}(h)$, however once a user joins her shard, she must stay online (and actively participates if she is a core member) until $\sigma_{pk_i}(h)$ expires. As a result, there does not exist any  explicit leave request. A leave  simply consists in not issuing a join request upon credential renewal.
    A consequence of this rule is that, in case user $i$ participates under credential $\sigma_{pk_i}(h)$ and spends her UTXO ($pk_i,sk_i$) before $\sigma_{pk_i}(h)$ expires, then $i$ continues to participate under $\sigma_{pk_i}(h)$ until $\sigma_{pk_i}(h)$ expires. Note that because a transaction only grants credentials after a delay, this rule does not allow a user to simultaneously own multiple credentials for the same stake. Note also that if $i$ is disconnected for a small amount of time this does not jeopardized the safety of the shard only its liveness. 
    \item Recall that the adversary has a bounded fraction $\mu$ of \emph{stake} in \nom. 
    To defend \nom against Sybil attacks (\ie, the fact that the adversary creates a considerable number of UTXOs with the objective of overpopulating each shard with malicious owners of those UTXOs), we require that each UTXO cannot be credited with more than $M$ stake, with $M$ some predefined constant.
    Consequently, by the fact that for any $h>0$ one credential $\sigma(h)$ represents exactly one UTXO, there is a bound $\mu_{cred} > \mu$ on the fraction of malicious credentials in \nom, which is reached when all malicious UTXOs have $1$ stake and all  honest ones maximize their stake, \ie, each honest UTXO has $M$ stake. Note that UTXOs with $M'$ stake, such that $M' > M$ may be handled by granting them $\lceil M'/M \lceil$ credentials, although we do not treat this case explicitly. Section~\ref{sec:security} analyzes the distribution of malicious credentials among shards.
\end{enumerate}
\noindent
Regarding the behavior of the adversary, there are a couple of remarks to note.
\begin{enumerate}
    \item At any time, the adversary might spend some selected UTXOs in order to create new ones and thus new credentials with the objective of targeting some shards. However, because of the initial $T$ blocks delay required to obtain the first credential for an UTXO (see Relation~\ref{eq:sojourn_time}), any newly created UTXO will give rise to a credential only after all existing credentials are renewed as well. Therefore, the adversary has no preferred strategy regarding transactions and forced renewal.
    \item Each block \(B_h\) contains a random seed, denoted by \(B_h.\rho\), which cannot, by construction, be either biased or predictable before the block is created (how such seeds are generated is detailed  in Section~\ref{sec:construction-next-block}). Thus by  Relation~\ref{eq:sojourn_time},   the adversary cannot determine  nor influence the value of renewed credentials. Consequently, for any blockchain height \(h\geq0\) and for any \(pk_i\), \(\sigma_{pk_i}(h+T)\) is unpredictable while for any \(0\leq h^\prime \leq h\), the sequence \((\sigma_{pk_i}(h^\prime+T))_{0\leq h^\prime\leq h}\) is computable and verifiable from the blockchain.
\end{enumerate}

\subsection{Shard Membership}
\label{sec:shard-membership}
As described above, during the period of time that elapses between the creation of an UTXO to its spending, the UTXO owner can participate to the blockchain construction by successively joining a series of shards.  In practice this may give rise to a  voluminous amount of join requests, which might be highly prejudicial to \nom's scalability and efficiency if each joining request led to the insertion of the newcomer in the core which run the distributed operations. Rather, by relying on PeerCube design (see Section~\ref{sec:set-of-ingredients}),  a newcomer joins the spare set of the shard and not its core set. This newcomer will be  a candidate for being elected as a member of the core set whenever  the core set will undergo a membership modification.  Management of the view composition, and election in the core set  is the purpose of the remaining of the section. 

\vspace{0.2cm}
\noindent{\textbf{View of a Shard} }
The view of a shard \(\mathcal{S}\) reflects the composition of both its core and spare sets, denoted respectively by \(\mathcal{S}_c\) and  \(\mathcal{S}_s\).  Update of the view is strongly correlated to blockchain events: any block appended to the blockchain is preceded, in each shard, by the update and the installation of the shard view. In the following, the view of shard $\mathcal S$ installed right before block $B_h$ is appended to the blockchain is denoted by \emph{view}$_{\mathcal S}(h)$. 
We have \(view_\mathcal{S}(h) = (\mathcal{S}_c(h), \mathcal{S}_s(h))\), where $\mathcal{S}_c(h)$ (resp. $\mathcal{S}_s(h)$) represent the composition of $\mathcal S$'s core set (resp. spare set)  at time $h$.

\vspace{0.2cm}
\noindent{\textbf{Update of the Shard View} }
When a newcomer (\ie a user under a valid credential) issues a request to join her shard $\mathcal S$, her request is propagated and broadcast to the members of \(\mathcal{S}_c\). 
Core members $i$ locally store the join  request in their buffer $b_i$ of pending requests. 
Note that expiration of credentials do not need to be locally memorized,  prior to being handled by the view update algorithm, since by Relation~\ref{eq:sojourn_time}, credentials can only expire when a new block is appended to the blockchain.  
Let $view_{\mathcal S}(h-1)$ be the current view of $\mathcal S$ when a (honest) core member $i \in \mathcal S_{c}(h-1)$ receives some valid  block $B_h$ (Section~\ref{sec:construction-next-block} details the creation of blocks). The following three steps are successively executed:
\begin{enumerate}
    \item A Byzantine vector agreement protocol is run among $\mathcal S_{c}(h-1)$ members  to decide on the set of newcomers:  core members $i$  propose their local buffer $b_i$, and the outcome of the protocol is a vector \(v(h)\) of newcomers such that non-null values for honest core members $i$ are equal to their buffer $b_i$. Each honest core member $i$ replaces its local buffer  $b_i$  with the union of the users of  the decided vector. We have \(b_i = \cup_{b_j \in v(h), b_j \neq \bot}b_j\).
    \item Each user $i \in \mathcal S_{c}(h-1)$ removes from $b_i$ the set \(r_\mathcal{S}(h)\) of users  whose credential expires with $B_h$.
    User $i$ initializes  a new spare set $S_{s}(h)$ with  $S_{s}(h) = b_i \cup S_{s}(h-1) \setminus r_\mathcal{S}(h)$, and  orders  $S_{s}(h)$. 
    \item  Each user $i \in \mathcal S_{c}(h-1)$ initializes a new core set $S_{c}(h)$  with  $S_{c}(h) = S_{c}(h-1) \setminus r_\mathcal{S}(h)$.
    If \(\mathcal{S}_c(h-1) \cap r_\mathcal{S}(h) \neq \emptyset\), some previous core members \(i \in \mathcal{S}_c(h-1)\) have credential that expire with \(B_h\). 
    As a consequence, an election among the users of \(\mathcal{S}_s(h)\) is carried out for \(i\)'s replacement, so as to keep \(|\mathcal{S}_c(h)| = \Smin\).
   The core election works as follows:
    \begin{enumerate}
        \item A random beacon protocol is run among $S_{c}(h-1)$ members to decide on a common random seed $\rho$. 
        \item A pseudo-random number generator $PRG(\rho)$  is initialized with $\rho$ as seed. 
	\item $PRG(\rho)$ is used to draw a random number \(j \in \llbracket 1, |S_{s}(h)|\rrbracket\). The \(j\)-th member of \(\mathcal{S}_s(h)\) is removed from \(\mathcal{S}_s(h)\) and added to \(\mathcal{S}_c(h)\). This process is repeated until \(|\mathcal{S}_c(h)| = \Smin\).
    \end{enumerate}
\end{enumerate}

Once these steps are completed, each core member $j$ installs  her new view $view^j_{\mathcal S}(h)$ with the new values of $S_{c}(h)$ and $S_{s}(h)$, signs it, and sends it to the spare members.
Once a spare receives $\mu_{core}\Smin+1$ signatures on the same view, it installs it.
In the meantime, each core member \(j\) resets its buffer \(b_j = \emptyset\).
Note that multiple join requests may lead a shard $\mathcal S$ to split into two shards, or, on the contrary, may lead two shards $\mathcal S^\prime$ and $\mathcal S^{\prime\prime}$ to merge within a single one $\mathcal S$. The treatment of such topological changes are omitted in the above procedure for space reasons, but can be derived from~\cite{ASLT11}. 
    
To summarize, the shard membership  procedure ensures that, for any shard $\mathcal S$ of \nom,  all  members of $\mathcal S$  install the same view $view_\mathcal{S}(h)$ before appending block $B_h$ to their copy of the blockchain.   

\vspace{0.2cm}
\noindent{\textbf{Diffusing Views} }
Merely installing the new view for each shard is not sufficient. We need the other shards of \nom to maintain this knowledge  to be able to verify any signed information exchanged during  inter-shard communication (\eg during the block proposal procedure, see Section~\ref{sec:construction-next-block}). 
Therefore, whenever a new view $view_\mathcal{S}(h)$ is installed along with its $\mu_{core}\Smin+1$ signatures, it is also broadcast to the whole network as a notification of the view update. 
Note that shards only store the last view $view_{\mathcal{S}'}(h)$ of any other shard $\mathcal{S}'$ and not the whole history of $\mathcal S'$ views. Moreover, a new view $view_{\mathcal{S}'}(h+1)$, can be verified against the last view $view_{\mathcal{S}'}(h)$, so that corrupted shards can only lie on their core members and omit newcomers.
%
\subsection{Construction of the Next Block of the Blockchain} 
\label{sec:construction-next-block}

We  propose a Byzantine resilient cross-shard mechanism to agree on a unique valid block, despite the presence of at most $f_{shard}$ corrupted shards (see Section~\ref{sec:security} for $f_{shard}$ computation). Indeed, the presence of an adaptive adversary may compromise the safety of some shards by succeeding in having more than a proportion $\mu_{core}$ of malicious users sitting in their core set. Although the probability of such event can be made arbitrarily low (see the analysis in Section~\ref{sec:security}), we must handle it.
The presence of corrupted shards put us in the same situation as in a consensus protocol: given the same initial chain, any shard is able to create the next block, and the decision must  be a unique block, despite malicious users lying or not responding.
As will be shortly described, agreeing on a unique valid block is efficiently and robustly achieved by running a Verifiable Byzantine Agreement among a subset of the shards of \nom randomly selected.

\vspace{0.2cm}
\noindent
\textbf{Reaching Consensus on the Next Block }
The process of creating a new block $B_h$ starts right after  $B_{h-1}$ has  been accepted. 
A committee of shards, denoted in the sequel by $\mathbb{C}$, is elected among the shards of \nom. The election of each of these shards relies on the seed of block $B_{h-1}$, derived from the random beacon protocol.
Once elected, committee $\mathbb{C}$ executes a verifiable Byzantine agreement to decide on the unique block $B_{h}$ to be appended to the blockchain. The main steps of this process are as follows:

\begin{enumerate}
    \item All shards $\mathcal S$ compute the elected committee $\mathbb{C}$, similarly to the core election procedure (see Section~\ref{sec:shard-membership}), i.e.,
    \begin{enumerate}
        \item Let $\mathbb{L}$ be the set of all the shards' labels (recall from Section~\ref{sec:shard-membership} that each shard diffuses its new view $view_\mathcal S(h)$). $\mathbb{L}$ is then ordered through a canonical order.
        \item A pseudo-random number generator $PRG(B_{h-1}.\rho)$ is initialized, where $B_{h-1}.\rho$ is the seed of the last block $B_{h-1}$.
        \item $PRG(B_{h-1}.\rho)$ is used to draw a random number \(j \in \llbracket 1, |\mathbb{L}|\rrbracket\). The \(j\)-th member of \(\mathbb{L}\) is removed from \(\mathbb{L}\) and added to \(\mathbb{C}\) (initially initialized to $\emptyset$). This process is repeated until \(\mathbb{C}\) contains $s_{\mathbb{C}}$ shards, with $s_{\mathbb{C}}=(f_{shard}/\mu_{corrupted})+1$. Recall that $f_{shard}$ is the maximal number of corrupted shards in \nom (whose computation is presented in Section~\ref{sec:security}), and $\mu_{corrupted}$ is the fraction of malicious nodes tolerated by the Verifiable Byzantine Agreement protocol (see Section~\ref{sec:set-of-ingredients}). 
    \end{enumerate}
    \item Members of committee $\mathbb{C}$  run the verifiable Byzantine Agreement  protocol, with their proposed block $B_{h}$ as input (the construction of the proposed block is described in the next paragraph). Finally the decision is a block $B_{h'}$ signed by $2 f_{shard}+1$ shards. 
    \item Block $b_h$ is broadcast in \nom and appended to  \nom users' copy of the blockchain. 
\end{enumerate}

\noindent
\emph{Security remark: } By definition of  $s_{\mathbb{C}}$, committee $\mathbb{C}$ cannot be corrupted, independently of the shards selected by the election. Committee $\mathbb{C}$ is still chosen randomly for two reasons. First, it naturally spreads the load of creating a block across the whole network. Second, it prevents corrupted shards from trying to manipulate the election process to get in the committee and slow it down. Note that at this stage a random seed is already available from the last block and thus there is no need to run a distributed random beacon. 

\noindent
\emph{Efficiency remark: } We rely on a leader-based Byzantine Agreement  algorithm to benefit from its optimistic efficiency. Indeed, since $f_{shard}$ can be made arbitrarily small (see Section~\ref{sec:security}), and the members of  committee $\mathbb{C}$ are randomly selected, we expect the first leader to almost always be an honest shard.

\vspace{0.2cm}
\noindent
\textbf{Construction of the Proposed Block. }
We finally describe how each shard $\mathcal S$  of $\mathbb{C}$ constructs its block  $B_h$  (see the above case 2). The construction results from an agreement  on the content of  block $B_h$ among the core members of $\mathcal S$ and on the generation of the seed of $B_h$.
Let $view_\mathcal{S}(h) = (\mathcal{S}_c(h), \mathcal{S}_s(h))$  be the current view of shard $\mathcal S$. 
\begin{enumerate}
    \item Each core member in $\mathcal{S}_c(h)$ proposes \emph{(i)} its list of pending transactions and \emph{(ii)} its VRF value seeded with $B_{h-1}.\rho$ together with the VRF proof, to the Byzantine Vector consensus protocol. The decision value is a vector of input values, such that non-null values for honest  core  members are  equal  to  their  list of pending transactions and their VRF value and VRF proof.
    
    \item Construction of block $B_h$ is then realized as follows.
    \begin{itemize}
        \item The hash of the previous block $B_{h-1}$ is inserted in $B_h$'s header.
        \item The union of transactions from the decided vector defines  $B_h$'s body.
        \item The hash of the concatenation of the VRF values of the decided vector defines the seed $B_h.\rho$ of $B_h$.
        \item The list of VRF proofs of the decided  vector is inserted in $B_h$'s header as a proof of randomness for  seed $B_h.\rho$.
    \end{itemize}
\end{enumerate}
The reason why the random beacon protocol is not reused is because it is supposed to be run within a non corrupted shard. For the We have different requirements. First, we  want the seed to be  close to random even in the case of corrupted shards. This does come at the cost of giving the adversary a bounded number of choices for the seed.
Second, we do not mind that a corrupted shard may decide to abort the computation of the seed, because we cannot prevent it from not proposing a block anyway.

\section{Security Analysis}
\label{sec:security}
We analyze the probability that some of the  shards of \nom are corrupted, that is that their core set contain more than $\mu_{core} \Smin$ malicious users. In the following we denote by $\nu$ the fraction of corrupted shards.  To conduct such an analysis, we  examine a simplified scenario. We approximate the behavior of \nom  by taking the amortized execution over one period of $T$ blocks. That is, we study the corruption probability when all the shards are built and the cores are elected over one period. This is equivalent to the scenario in which  all credentials are synchronously renewed at the same block. Note that, for a fixed number of active users, the number of credential renewals, core election, and topological changes is statistically the same for every period of length $T$.

\subsection{Corruption Probability of a Core Set during a Period of  $T$ Blocks}

Let 
$s$ be the  size of shard $\mathcal S$, 
$\mu_{shard}$ be a bound on the ratio of malicious users within  $\mathcal S$, and $\mu_{core}$ be the fractional resiliency of both the  Vector Agreement protocol and Random Beacon one. We assume that \(0\leq \mu_{shard} < \mu_{core} \leq \mu\). 
We compute an upper bound on the probability that the fraction of malicious users in the core set is higher than $\mu_{core}$ by the end of the period.
As described in Section~\ref{sec:shard-membership}, the core set is elected by randomly taking $\Smin$ credentials from shard \(\mathcal{S}\), without replacement. 
Let $Y$ be the random variable equal to the number of malicious credentials within the core, \ie, $Y$ follows an hypergeometric distribution whose probability mass function  is given by
\begin{equation}\label{eq:pmf}
     \forall k\in\llbracket0,\Smin\rrbracket, \proba{Y = k} = \binom{\lfloor s\mu_{shard} \rfloor}{k} \binom{\lfloor s(1-\mu_{shard}) \rfloor}{\Smin-k}\binom{s}{\Smin}^{-1}. 
\end{equation}
We are interested in deriving the probability that  after \(T\) core renewals the core set \(\mathcal{S}\) is corrupted. 
The core set corruption refers to the situation where the proportion of malicious credentials in the core exceeds \(\mu_{core}\).
Applying the Hoeffding bound~\cite{hoeffdingbound} on Relation (\ref{eq:pmf}) leads to the following bound
\begin{align*}
    \proba{Y/\Smin \geq \mu_{core}} \leq e^{-2(\mu_{core}-\mu_{shard})^2 \Smin}. 
\end{align*}
Thus, assuming that the fraction of malicious users in a shard is below $\mu_{shard}$, the corruption probability over $T$ blocks exponentially decreases when $\Smin$ increases.

\subsection{Distribution of Malicious Credentials among all Shards}
The above section assumes that the fraction of malicious users in all the  shards is below $\mu_{shard}$. In this section we compute an upper bound on the probability that this assumption does not hold.
We make simplification assumptions on how the shards are formed. First, we assume that  there are $K$ shards of size $S$, giving rise to \ie $N:=SK$ credentials in total. Second, we assume that shards configuration in \nom during the concerned period results from a random credential assignment to all the shards.
Recall that $\mu_{cred}$ is the overall ratio of malicious credentials. 
Let $X_i$ be the random variable representing  the number of malicious credentials in the $i$-th shard, with $1<i<K$.
And finally, we note \(\mathbf{X} = (X_1, \dots, X_k) \in \{0, S\}^K\) be the  vector made of these $K$ random variables.  Random variable \(\mathbf{X}\) represents the distribution of malicious credentials in \nom. It  follows a multivariate hypergeometric distribution, \ie,  each of the $N = S K$ credentials is assigned to a shard.  We analyse  the shard  assignment of a random sample of size $N\mu_{cred}$.
Let  \(I\) be  the set of vectors representing \nom when \(N\mu_{cred}\) credentials are malicious. We have

 $$  \mathrm{I} = \{\mathbf{x}\in[0,S]^K \mid \sum_{i=1}^K x_i = N\mu_{cred} \}$$ \t and
   $$ \forall \mathbf{x}\in \mathrm{I}, \proba{\mathbf{X} = \mathbf{x}} = \binom{N}{N\mu_{cred}}^{-1} \prod_{i=1}^{K} \binom{S}{x_i}.$$

We are interested in computing the probability that a given shard $j$ among the $K$ ones  contains more than \(m\) malicious credentials, that is, let  $\mathrm{I}_{m,j}$ be defined as follows $$\mathrm{I}_{m,j}  = \{ \mathbf{x} \in \mathrm{I} \mid x_j \geq m \}.$$
We have:
\begin{eqnarray*}
    \proba{X_j\geq m} & = &\proba{\mathbf{X} \in \mathrm{I}_{m,j}} \\ 
    & =& \sum_{x\in\mathrm{I}_{m,j}}  \binom{N}{N\mu_{cred}}^{-1} \prod_{i=1}^{K} \binom{S}{x_i}\\
    & = & \sum_{k=m}^{S} \binom{S}{k} \binom{N}{N\mu_{cred}}^{-1} \sum_{\substack{x_1,\dots,x_{K-1}\in[0,S] \\ \sum_{1\leq i\leq K-1}x_i = N\mu_{cred}-k}} \prod_{1\leq i \leq K, i\neq j} \binom{S}{x_i}.
\end{eqnarray*}

Knowing that $\sum_{1\leq i \leq K-1} x_i = N\mu_{cred}-k$ and $\sum_{1\leq i \leq K-1} S = N-S$, we can apply Vandermonde's identity:
\begin{align*}
    \forall j, \proba{\mathbf{X} \in \mathrm{I}_{m,j}} = \sum_{k=m}^{S} \binom{N}{N\mu_{cred}}^{-1} \binom{S}{k}\binom{N-S}{N\mu_{cred}-k}.
\end{align*}
We now get our result by applying first the (univariate) Hoeffding bound, and then the union bound.
\begin{align*}
    &\forall j, \proba{\mathbf{X} \in\mathrm{I}_{S\mu_{shard},j}} \leq e^{-2(\mu_{shard}-\mu_{cred})^2S}.
\end{align*}
Thus  the probability that at least one shard of the system contains more than \(\mu_{shard}S\) malicious credentials is bounded by
\begin{eqnarray*}
      \proba{\mathbf{X} \in\cup_{j=1}^K\mathrm{I}_{S\mu_{shard},j}} &\leq& Ke^{-2(\mu_{shard}-\mu_{cred})^2S} \\
    & = & e^{-(2(\mu_{shard}-\mu_{cred})^2S - \ln{K})}. 
\end{eqnarray*}
Term $\cup_{j=1}^K\mathrm{I}_{S\mu_{shard},j}$ is the set of shards assignations to malicious credentials, such that at least one shard has a fraction greater than or equal to $\mu_{shard}$ of malicious credentials.
Moreover, due to the union bound, this upper bound also holds if the shards  have different sizes and $S$ is the minimum, hence, we can simply use $S:= \Smin$. As for $K$, the worst case is reached  when there is a maximal number of shards, \ie $K:=N/\Smin$.

\subsection{Putting it all Together}

In the previous subsection we got exponentially decreasing bounds on the probability that at least one shard is corrupted, \ie, proving security when the bound on the number of malicious shards $f_{shard}$ is set to $0$. We let for future work the generalization of this calculation with arbitrary values of $f_{shard}$, which would give us tighter parameters.

The adversary has a fraction $\mu$ of stake. Requiring each credential to be associated to at most $M$ stake gives us the following (worst case) ratio of malicious credentials, which is reached when each  malicious UTXOs has $1$ stake and each  honest one maximizes its stake, \ie, has $M$ stake. We then have: $$
\mu_{cred} =  \frac{1}{1+M^{-1}(\mu^{-1}-1)}.
$$

Thus $M$ should be as small as possible to decrease the adversary effective stake. However low values of $M$ may require users to participate with  a large number of credentials in parallel, increasing the communication cost for individual users.
Knowing $\mu$ and security parameter $\kappa$, the parameters $\mu_{shard}$ and $\Smin$ can be obtained by solving the following inequalities 
\begin{align*}
 &\mu_{shard} \leq \mu_{cred} + \sqrt{\frac{\kappa-\ln{\frac{N}{\Smin}}}{2 \Smin}} 
 \;\;\;\;\; \mbox{and}  \;\;\;\;\;
  \Smin \geq \frac{\kappa}{2(\mu_{core}-\mu_{shard})^2}.
\end{align*}

\section{Conclusion \& future work}
\label{sec:conclusion}

In this paper we have presented \nom a new blockchain protocol which aims at improving scalability of the block-wise Byzantine agreement approach by combining sharding techniques, users presence and stake transfer to operate in a PoS setting. Each block at height $h$ in the blockchain is by design  unique (no fork), and once a block is accepted in the blockchain, the next one is created by a sub-committee of shards whose selection depends on the random seed of the last accepted block. 

The next step is to take into account the stake associated with each credential as weights into both the core election and the election of the shard in charge of creating the next block. This will allow us to get rid of the $\mu_{cred}-\mu$ gain in adversarial power, while keeping the remaining of the security arguments similar. More generally, refinements of the security analysis will give us the ability to instantiate \nom with better parameters while keeping the same security level.

We also plan to implement a prototype of \nom to demonstrate its efficiency and scalability properties, and to showcase some possible applications.

\section{Acknowledgements}
We are thankful to Gérard Memmi (LTCI Telecom ParisTech), and David Leporini, Guillaume Hebert and Thomas Domingos (Atos BDS) for their help and fruitful discussions. This work was carried as part of the Blockchain Advanced Research \& Technologies (BART) Initiative and the Institute for Technological Research SystemX, and therefore granted with public funds within the scope of the French Program \textit{Investissements d'Avenir}.

\balance
\bibliographystyle{splncs04}

\bibliography{blockchain-2018}
\end{document}